\RequirePackage{etoolbox}
\newtoggle{daconf}

\iftoggle{daconf}{%
  \documentclass[12pt,conference,compsoc,onecolumn]{drivingassessment}
  \usepackage[round,sort,authoryear]{natbib}
  \bibliographystyle{abbrvnat}
  \setcitestyle{authoryear}
  \setlength{\bibsep}{6pt plus 0.3ex}
}{%
  \documentclass[conference]{IEEEtran}
  \usepackage[square,sort,numbers]{natbib}
  \bibliographystyle{abbrvnat}
}

\usepackage[breaklinks,colorlinks,bookmarksopen,bookmarksnumbered]{hyperref}
\hypersetup{
    colorlinks=true,
    linkcolor=black,
    filecolor=black,      
    urlcolor=black,
    citecolor=black,
}
\RequirePackage{snapshot}

\usepackage{subcaption}
\usepackage{capt-of}
\usepackage{balance}

\usepackage{cuted}

\usepackage[rightcaption]{sidecap}
\usepackage{booktabs}
\usepackage{graphicx}
\usepackage{float}
\usepackage{enumitem}

\usepackage{wrapfig}
\usepackage{mdframed}

\usepackage{moreverb}
\usepackage{verbatim}

\usepackage{color}

\usepackage{parskip}

\usepackage{latexsym}

\usepackage[sharp]{easylist}

\usepackage{multirow}

\usepackage{inconsolata}
\newcommand{\weblink}[1]{\path{#1}}

\usepackage[usenames,dvipsnames,table]{xcolor}

\usepackage{amsmath}
\usepackage{amsfonts}
\usepackage{amssymb}
\usepackage{amsthm}

\usepackage{algorithmic}
\usepackage{algorithm}
\usepackage{cases}

\renewcommand{\eqref}[1]{(\ref{eq:#1})}

\newcommand{\figref}[1]{Fig.~\ref{fig:#1}}
\newcommand{\tabref}[1]{Table~\ref{tab:#1}}

\newcommand{\traj}[1]{\uppercase{#1}}
\newcommand{\tgroup}[1]{\uppercase{\textit{#1}}}

\newcommand{\additionStart}{\begin{mdframed}[hidealllines=true,backgroundcolor=yellow!20]}
\newcommand{\additionEnd}{\end{mdframed}}

\definecolor{tred}{RGB}{205,92,92}
\definecolor{tgreen}{RGB}{60,179,113}
\definecolor{tgrey}{RGB}{169,169,169}
\definecolor{tdarkblue}{RGB}{0,0,255}
\definecolor{tlightblue}{RGB}{30,144,255}
\definecolor{tpurple}{RGB}{138,43,226}

\newcommand{\tred}{\textcolor{tred}{\textbf{red}}}
\newcommand{\tgreen}{\textcolor{tgreen}{\textbf{green}}}
\newcommand{\tgrey}{\textcolor{tgrey}{\textbf{grey}}}
\newcommand{\tdarkblue}{\textcolor{tdarkblue}{\textbf{dark blue}}}
\newcommand{\tlightblue}{\textcolor{tlightblue}{\textbf{light blue}}}
\newcommand{\tpurple}{\textcolor{tpurple}{\textbf{purple}}}

\usepackage{fancyhdr}
\fancypagestyle{firststyle}
{
  \fancyhf{}
  
   \fancyfoot[L]{
     * Corresponding author: \texttt{fridman@mit.edu}
   }
}

\begin{document}

\title{Hacking Nonverbal Communication Between Pedestrians and Vehicles in Virtual Reality}
  
\newcommand{\authorspace}{\hspace{0.2in}}
\author{
Henri Schmidt \authorspace
Jack Terwilliger \authorspace
Dina AlAdawy \authorspace
Lex Fridman*\\
Massachusetts Institute of Technology\\
}


\maketitle

\captionsetup[sub]{labelformat=parens}
\begin{strip}
  \centering
  \captionsetup{type=figure}
  \begin{subfigure}[b]{0.43\textwidth}
    \includegraphics[width=\textwidth]{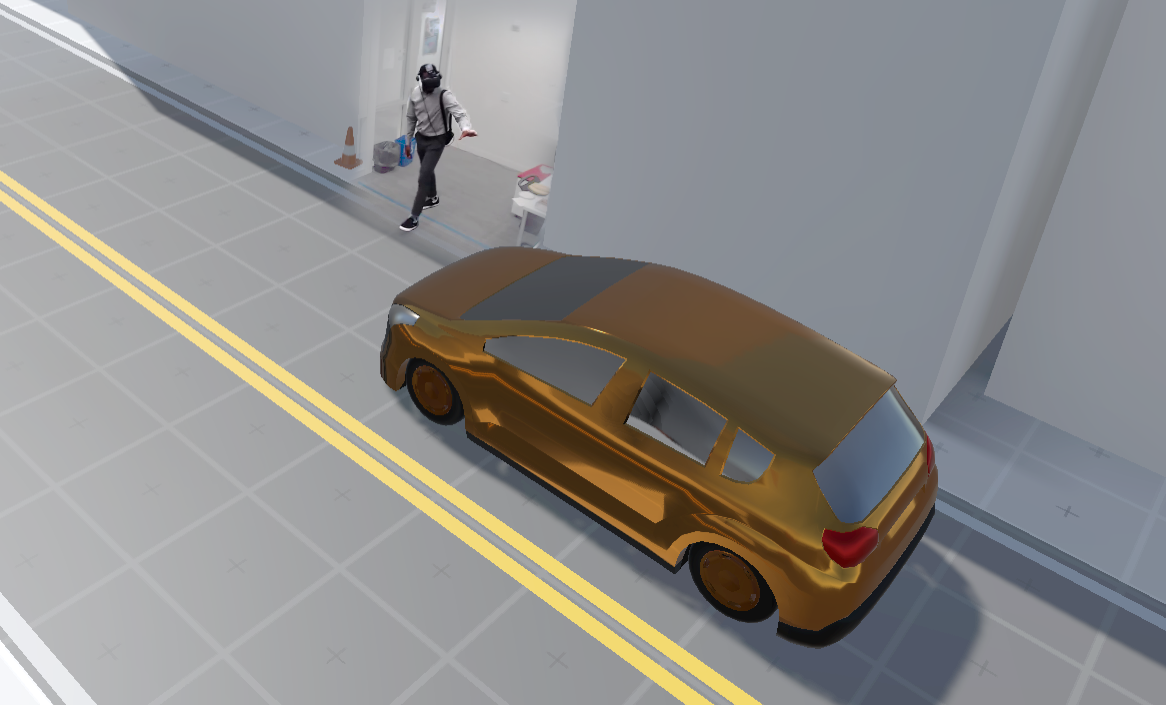}
    \caption{Participant crossing the virtual street}
    \label{fig:walker}
  \end{subfigure}\hspace{0.15in}
  \begin{subfigure}[b]{0.16\textwidth}
    \includegraphics[width=\textwidth]{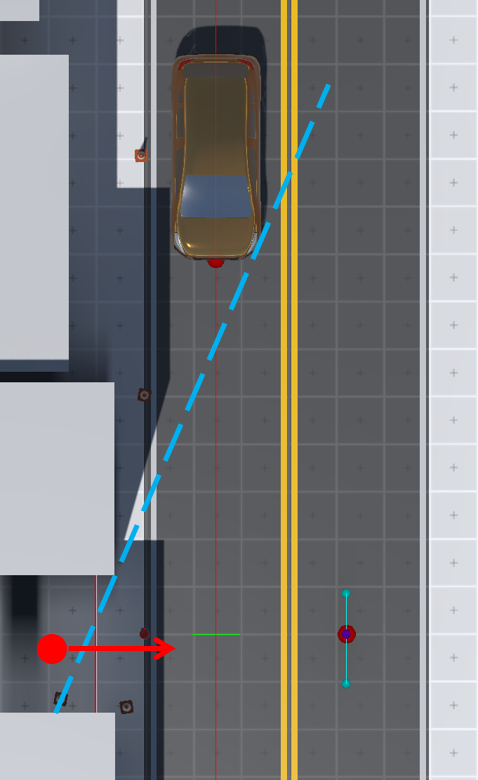}
    \caption{Top view diagram}
    \label{fig:topdown}
  \end{subfigure}\hspace{0.15in}
  \begin{subfigure}[b]{0.348\textwidth}
    \includegraphics[width=\textwidth]{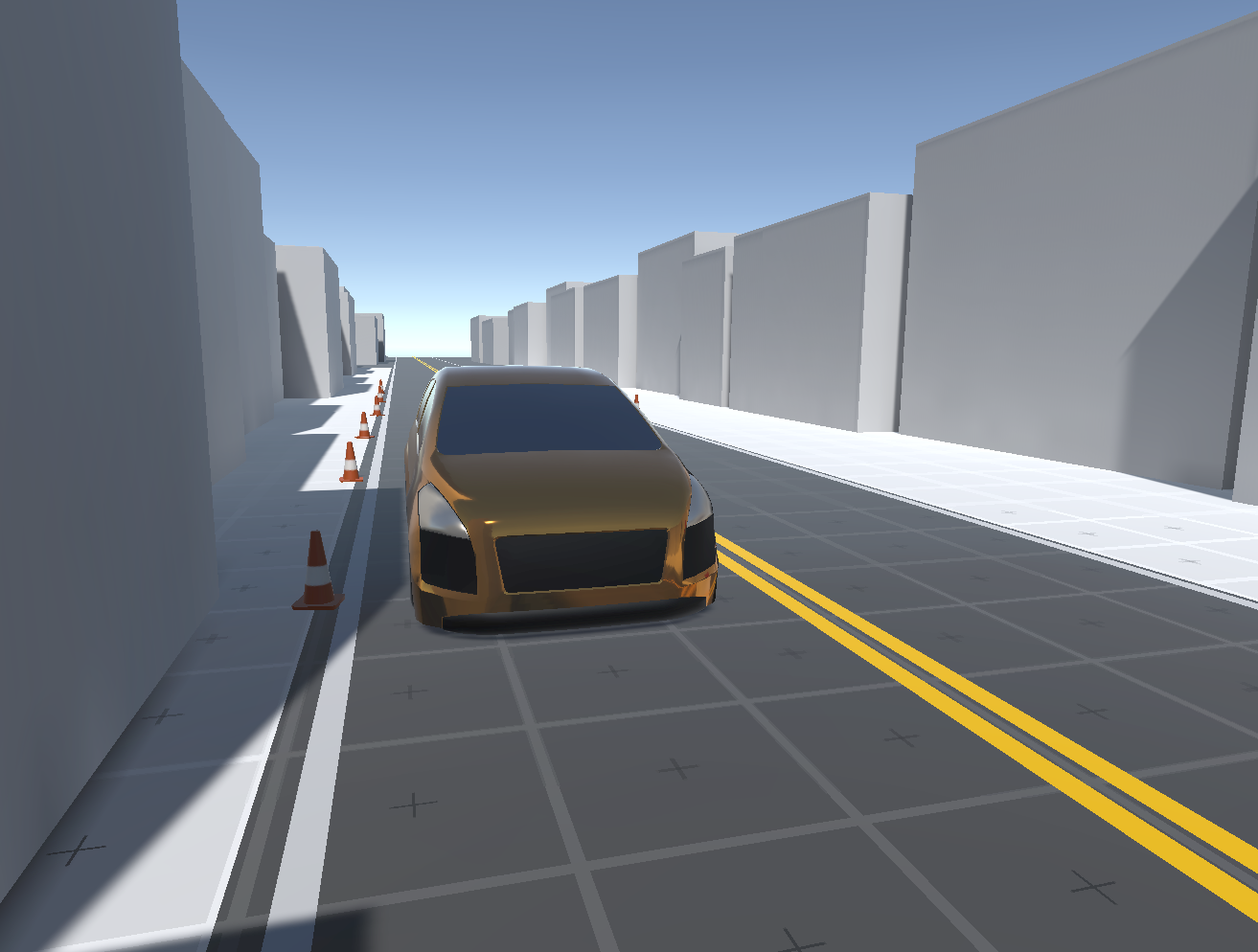}
    \caption{First person perspective view}
    \label{fig:fpv}
  \end{subfigure}
  \caption{Visualization from various perspectives of the pedestrian crossing scenario in virtual
    reality. \figref{walker} visualizes the crossing experience. \figref{topdown} shows a top down view of the
    intersection where the red dot indicates starting position and the blue dashed line indicates the line of sight
    threshold where the vehicle becomes visible. \figref{fpv} provides the first person view of the subject in virtual
    reality based on the orientation of their head.}
  \label{fig:illustrations}
\end{strip}

\begin{abstract}%
We use an immersive virtual reality environment to explore the intricate social cues that underlie non-verbal communication involved in a pedestrian's crossing decision. We ``hack'' non-verbal communication between pedestrian and vehicle by engineering a set of 15 vehicle trajectories, some of which follow social conventions and some that break them. By subverting social expectations of vehicle behavior we show that pedestrians may use vehicle kinematics to infer social intentions and not merely as the state of a moving object. We investigate human behavior in this virtual world by conducting a study of 22 subjects, with each subject experiencing and responding to each of the trajectories by moving their body, legs, arms, and head in both the physical and the virtual world. Both quantitative and qualitative responses are collected and analyzed, showing that, in fact, social cues can be engineered through vehicle trajectory manipulation. In addition, we demonstrate that immersive virtual worlds which allow the pedestrian to move around freely, provide a powerful way to understand both the mechanisms of human perception and the social signaling involved in pedestrian-vehicle interaction.
\end{abstract}

\thispagestyle{firststyle}
\setlength{\footskip}{20pt}

\section{Introduction}\label{sec:introduction}

One of the challenging aspects in the design of autonomous vehicles is their communication with other, non-autonomous
participants in traffic. Specifically the interaction with pedestrians requires clear communication of intent to allow
for safe interactions \citep{rasouli2017agreeing}. If autonomous vehicles will be more prevalent in the future,
yielding to pedestrians under all circumstances (i.e. conservative driving behavior) may no longer be feasible as an
interaction strategy.  It has been shown that communicating the intention not to yield to pedestrians in certain traffic
situations can significantly increase traffic flow \citep{gupta2018negotiation}.  Finding ways to communicate such
intentions to pedestrians in a way that is easy to understand and assertive but \textit{safe} for the pedestrian remains
an open challenge of autonomous driving.  In this paper we investigate how vehicle kinematics can be ``hacked'' to
project intent and manufacture non-verbal communication cues that are actionable and interpretable by the interacting
pedestrian.

\section{Related Work}\label{sec:related_work}

Pedestrian-vehicle-interactions in the form of road crossings have thus far mostly been studied  as a problem of gap size and time to arrival, among the methods used are two-dimensional as well as curved screens \citep{oxley2005crossing}, announcing crossing intent while observing actual intersections \citep{schwebel2008validation} and immersive Virtual Reality (VR) \citep{clancy2006road, simpson2003investigation}. While these studies do of course consider vehicle movement, it is taken in a physical context and explored in terms of remaining distance or time for the pedestrian to reach the other side of the road. 

Current research regarding the general interaction between \emph{autonomous} vehicles and pedestrians has been focused on external Human Machine Interfaces (EHMIs). These concepts revolve around variations of displays, lights or projections placed inside or outside of the vehicle \citep{mahadevan2018communicating, clamann2017evaluation, risto2017human, deb2018investigating, dey2018interface}. Such mechanisms are intended to replace explicit gestures from the driver towards pedestrians intending to cross \citep{mahadevan2018communicating, risto2017human}. Such mechanisms have previously also been studied using virtual reality \citep{deb2018investigating}.

As EHMIs are a novel concept in driver pedestrian interactions they bring with them various issues and design challenges which have yet to be overcome. Such challenges include for instance the design of interfaces which are discernable at the distance of an approaching vehicle \citep{clamann2017evaluation}, as well as visible and understandable in the context of busy intersections \citep{risto2017human}. In addition, the extend to which the driver cues they are intended to replace actually aide in pedestrian-vehicle interactions as they occur today is questionable \citep{dey2017pedestrian, rothenbucher2016ghost}.

Our work aims to explore vehicle kinematics as an alternative form of vehicle pedestrian communication under special consideration of Autonomous Vehicles (AVs).

It has ben shown that the way non-humanoid robots use shared space in a ``passive'' or ``assertive'' manner when interacting with humans is perceived as giving social cues conveying the ``emotional state'' and consequently the intentions of said robots. This holds true ``regardless of whether that robot is capable of having emotional states or not.''\citep{fiore2013toward}

The role of vehicle kinematics in particular as a means of social communication has previously been studied by means observation, for instance the concept of ``motion in context'' in \citep{risto2017human}, as well as the importance of ``motion patterns and vehicle behavior'' as observed in \citep{dey2017pedestrian}. Specifically interactions between pedestrians and (seemingly) autonomous vehicles have been investigated as a Wizard-of-Oz study in absence of EHMIs \citep{rothenbucher2016ghost}.

As is apparent form the previous paragraphs, virtual reality has already been established as a tool for studying vehicle-pedestrian-interactions \citep{clancy2006road, simpson2003investigation, deb2018investigating, bhagavathula2018reality}. VR is successfully used in various various fields, including psychology and visual perception experiments \citep{wilson2015use}. The use of screen based, two dimensional virtual interactions for studying pedestrian interactions in particular has been validated by multiple studies \citep{oxley2005crossing, schwebel2008validation}. While an objective measurement of immersion (the overall realism and fidelity of a virtual environment) is difficult, it has been established that increased immersion is a desireable trait in experiment design \citep{wilson2015use} and beneficial towards the spacial understanding of the simulated environment \citep{bowman2007virtual}. Such effects are aided by stereoscopic rendering (providing a distinct image to each eye of the VR user, allowing for life like depth perception), head tracking (translating the visual virtual perception according to the actual head movements of the VR user) and a large field of regard (FOR) (the overall size of the visual field a VR user can cover by means of head movement) \citep{bowman2007virtual}.
While some previous studies have found that the the scale of virtual worlds is not always perceived correctly, it has been shown that this effect can be mitigated if participants are allowed to traverse such environments on foot \citep{wilson2015use, kelly2013more}.
A investigation into the applicability of immersive virtual reality for studying road crossing decisions based on time to arrival found that, while there are differences in the estimated vehicle speed between real-world and virtual scenarios, these did not have a measurable effect on pedestrian crossing decisions \citep{bhagavathula2018reality}.



 \begin{table*}[!ht]
  \caption{Trajectories}
  \label{tab:trajectories}
  \centering\resizebox{1\textwidth}{!}{%
\begin{tabular}{lrrllp{10cm}}
\toprule
          Trajectory &  Dist. (m) &  TTA (s) &    Velocity &       Group &                                                                                                                                                                                                                                                                                                                              Description \\
\midrule
  deterrent\_50kph\_2s &      27.78 &      2.0 &    constant &   DETERRENT &                                                                                                                                                                                                                                                                   Constant speed, low tta. Intended to deter participants from crossing. \\
  deterrent\_40kph\_4s &      11.11 &      4.0 &    constant &   DETERRENT &                                                                                                                                                                                                                                                                   Constant speed, low tta. Intended to deter participants from crossing. \\
    rolling\_yield\_5m &      14.58 &      9.0 &  decelerate &       YIELD &                                                                                                                                                                                                                        Deceleration from 20 km/h to 3 km/h in 3s, deceleration completes 5m from the intersection with 6s remaining tta. \\
    rolling\_yield\_8m &      18.75 &      9.0 &  decelerate &       YIELD &                                                                                                                                                                                                                        Deceleration from 20 km/h to 3 km/h in 3s, deceleration completes 8m from the intersection with 6s remaining tta. \\
 15kph\_acceleration  &      27.50 &      8.0 &  accelerate &  15\_KPH\_SET &                                                                                                                                                                                                                                                                                    The vehicle accelerates from 1 km/h to 15 km/h in 3s. \\
 15kph\_deceleration  &      45.83 &      8.0 &  decelerate &  15\_KPH\_SET &                                                                                                                                                                                                                                                                                   The vehicle decelerates from 45 km/h to 15 km/h in 3s. \\
 15kph\_uniform\_speed &      33.33 &      8.0 &    constant &  15\_KPH\_SET &                                                                                                                                                                                                                                                                   The vehicle approaches at a constant speed of 15 km/h with a tta of 8s \\
 40kph\_deceleration  &     162.67 &      8.0 &  decelerate &  40\_KPH\_SET &                                                                                                                                                                                                                                                                       The vehicle decelerates from 106.4 km/h to 40 km/h over 8 seconds. \\
 40kph\_acceleration  &      61.11 &      8.0 &  accelerate &  40\_KPH\_SET &                                                                                                                                                                                                                                                                      The Vehicle will accelerate from 15 km/h to 40 km/h over 8 seconds. \\
 40kph\_uniform\_speed &      88.89 &      8.0 &    constant &  40\_KPH\_SET &                                                                                                                                                                                                                                                                  The vehicle drives by at a constant speed of 15 km/h with a tta of 8 s. \\
   breaking\_on\_enter &      12.00 &      4.8 &       other &       OTHER &                                                                                                                                                                                                                                                       If the pedestrian enters the lane the vehicle decelerates 1.8 km/h with -1.3 m/s\textasciicircum 2 \\
   conf\_jump\_rolling &      12.00 &     15.0 &       other &  SUBVERSION &  The vehicle moves at a constant, slow pace. Looking at the vehicle or stepping short of 0.8 m from the curb will cause the vehicle to accelerate from 0.8 km/h to 3 km/h with 3.5 m/s\textasciicircum 2 and then immediately decelerate back 0.8 m/s\textasciicircum 2. This is repeated if the participant takes their gaze of the vehicle and then looks at it again. \\
   conf\_jump\_stopped &       6.00 &      NaN &       other &  SUBVERSION &     The vehicle is stopped 6 m from the intersection. Looking at the vehicle or stepping short of 0.8 m from the curb will cause the vehicle to accelerate from to 3 km/h with 3.5 m/s\textasciicircum 2 and then immediately decelerate back to a stop. This is repeated if the participant takes their gaze of the vehicle and then looks at it again. \\
  conf\_malicious\_acc &      20.00 &      NaN &       other &  SUBVERSION &                                                  The vehicle starts of moving steadily at 2 km/h from 20 m distance, which leads to a perceived tta of 36s. If the pedestrian is in the lane of travel and not looking at the vehicle it will accelerate with 3.5 m/s\textasciicircum 2  to 8 km/h. This trajectory was designed to be openly malicious. \\
  conf\_distance\_mirr &        NaN &      NaN &       other &  SUBVERSION &        The vehicle mirrors the movements of the pedestrian. It will take the rolling average of the pedestrians position over 0.66s with a delay of 1.8s and position itself at twice the pedestrians distance from the point of intersection at that time. The vehicle ``mirrors'' the actions of the participant with a slight delay.  \\
\bottomrule
\end{tabular}
}\end{table*}

\section{Methods}\label{sec:methods}

\begin{figure}
  \begin{center}
      \includegraphics[width=0.8\columnwidth]{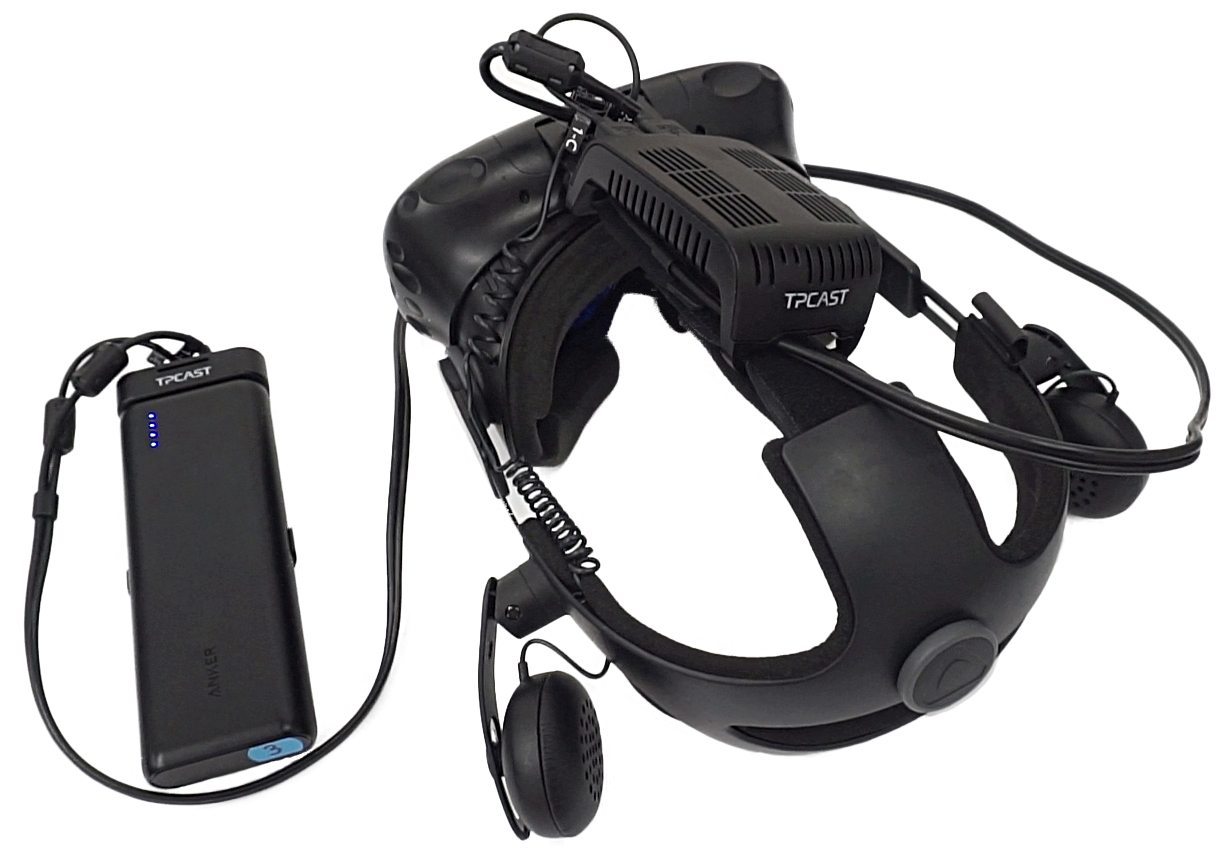}
      \caption{HTC Vive Virtual Reality Headset with TPCast Wireless Transceiver}
      \label{fig:vive}    
  \end{center}
\end{figure}

\begin{figure*} 
  \centering
  \begin{subfigure}[b]{0.33\paperwidth}
    \includegraphics[width=0.33\paperwidth]{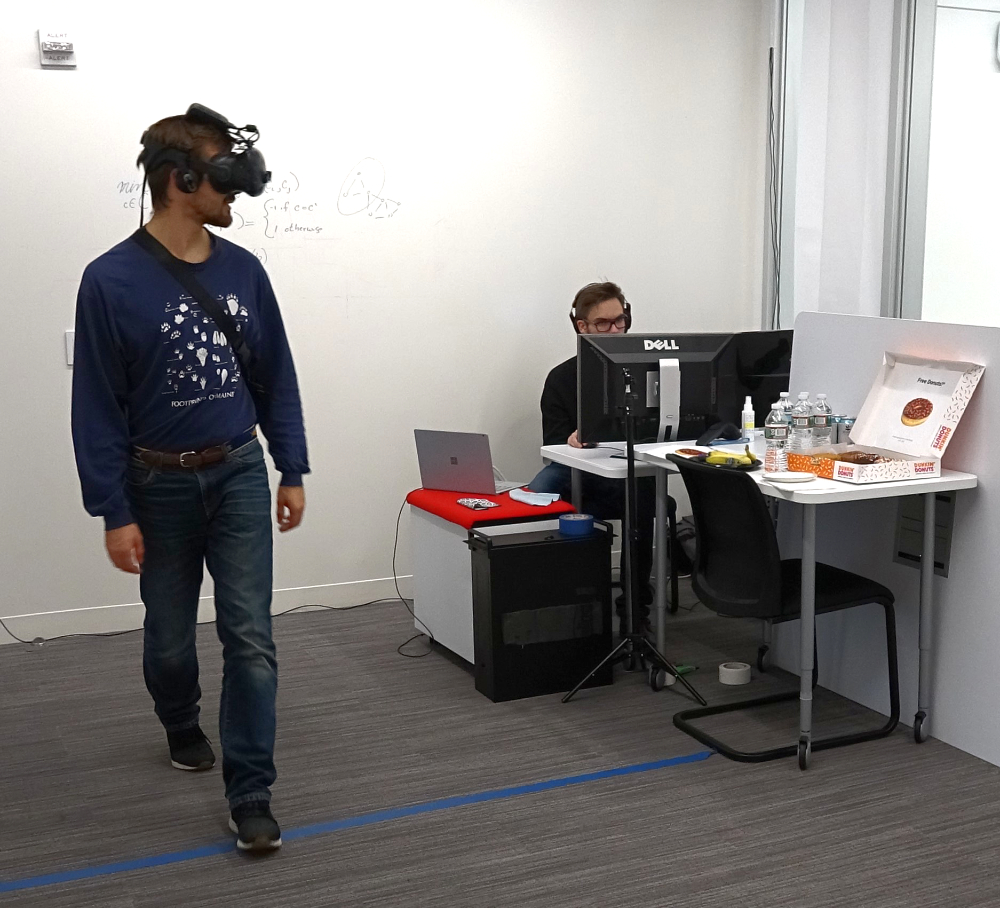}
    \caption{Participant with HMD stepping into virtual street}
    \label{fig:setup_photo}
    \end{subfigure}
  \hspace{0.01in}
  \begin{subfigure}[b]{0.47\paperwidth}
    \includegraphics[width=0.47\paperwidth]{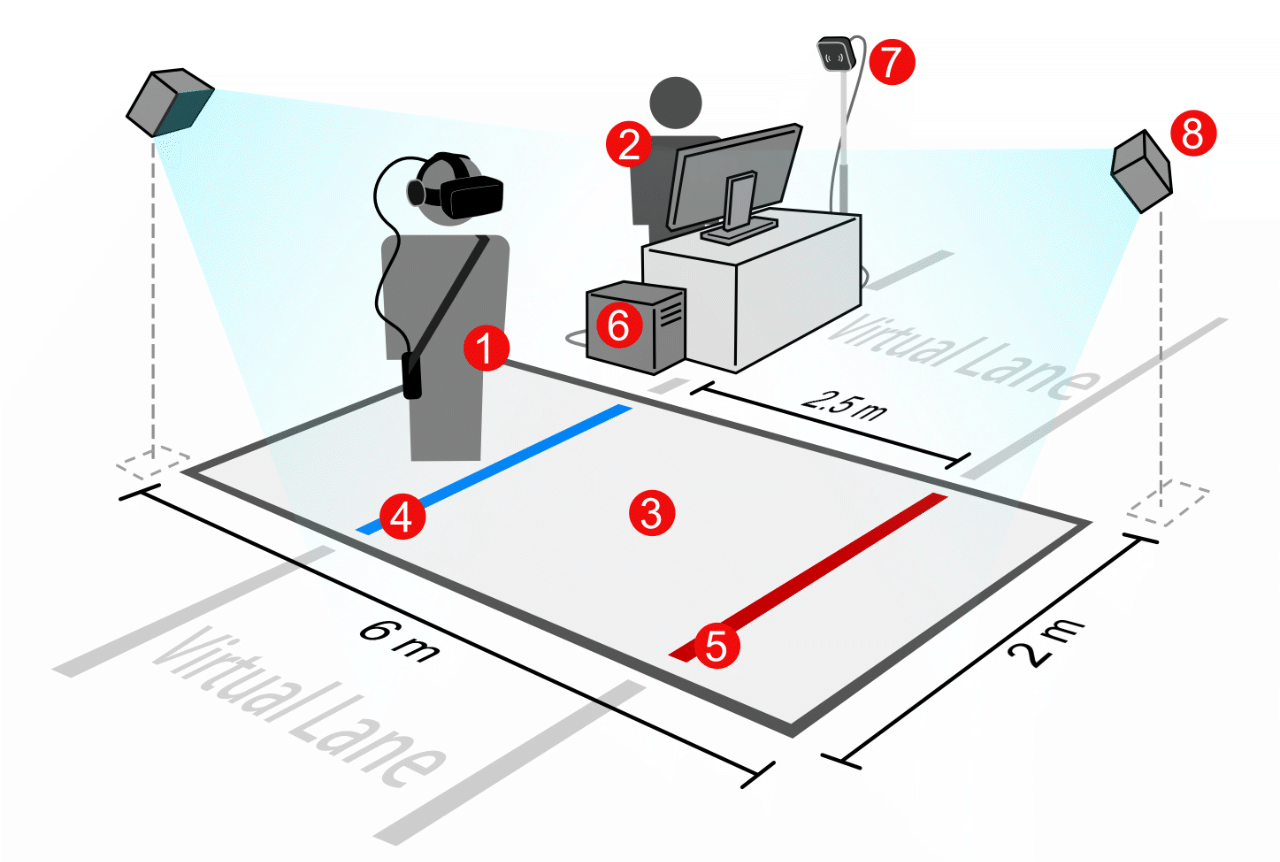}
    \caption{Diagram of the experimental setup}
    \label{fig:setup_illustration}
  \end{subfigure}
  \caption{Experimental Setup: (1) Participant with HMD, Wireless Transceiver and Battery, (2) Experimenter, (3)
    Walkable VR space (6 m x 2 m), (4) Virtual Curb, (5) End of Lane, (6) Simulation Computer, (7) VR Transceiver, (8)
    Tracking Base-Station}
  \label{fig:setup}
\end{figure*}

To understand the potential for social cues in vehicle kinematics, we studied the reaction of pedestrians towards vehicles exhibiting different kinds of behaviors in a road crossing situation.

We engineered these behaviors to juxtapose interactions which comply with what we expected to be the social convention of such interactions (with reference also to \citep{risto2017human}) with behaviors which would be unexpected. For both scenarios the time available to the pedestrian to cross is kept identical.

Observing a difference in reaction between the regular and subversive vehicle behaviors would then allow us to conclude that participants derive cues towards the intentions of the vehicle from the vehicle kinematics, as our testing environment features no other means of communication from the vehicle. 

``Social cues'' in the context of this paper refer to the presence of information aiding pedestrians in inferring the current and future behavior of the vehicle, beyond the pure physicality of the executed movement. The question is if pedestrians view the movement of the vehicle simply as a function of distance over time or as decisions of an intelligent entity whose goals need to be aligned with their own.


To study these interactions between pedestrians and vehicles we created an immersive virtual reality
environment:

\subsection{Setup and Virtual Environment}\label{sec:setup}

Virtual reality offers multiple benefits in this situation: It allows us to explore edge cases in human
vehicle communication without any risk to our human participants in cases where the communication fails. The simulation inside a virtual environment further provides precise experimental control over the vehicle movements and repeatability of scenarios across participants, as well as precise data-recording mechanisms. 

We created our virtual reality setup using the \textit{Unity3D} game engine, which allows for quick prototyping and easy
integration of virtual reality. The \textit{Head Mounted Display (HMD)} we chose for this experiment is the \textit{HTC Vive} (\figref{vive}). The tracking of the HMD allows the participant to traverse our virtual environment on foot with a natural range
of motion. 

\begin{figure}
  \begin{center}
      \includegraphics[width=0.9\columnwidth]{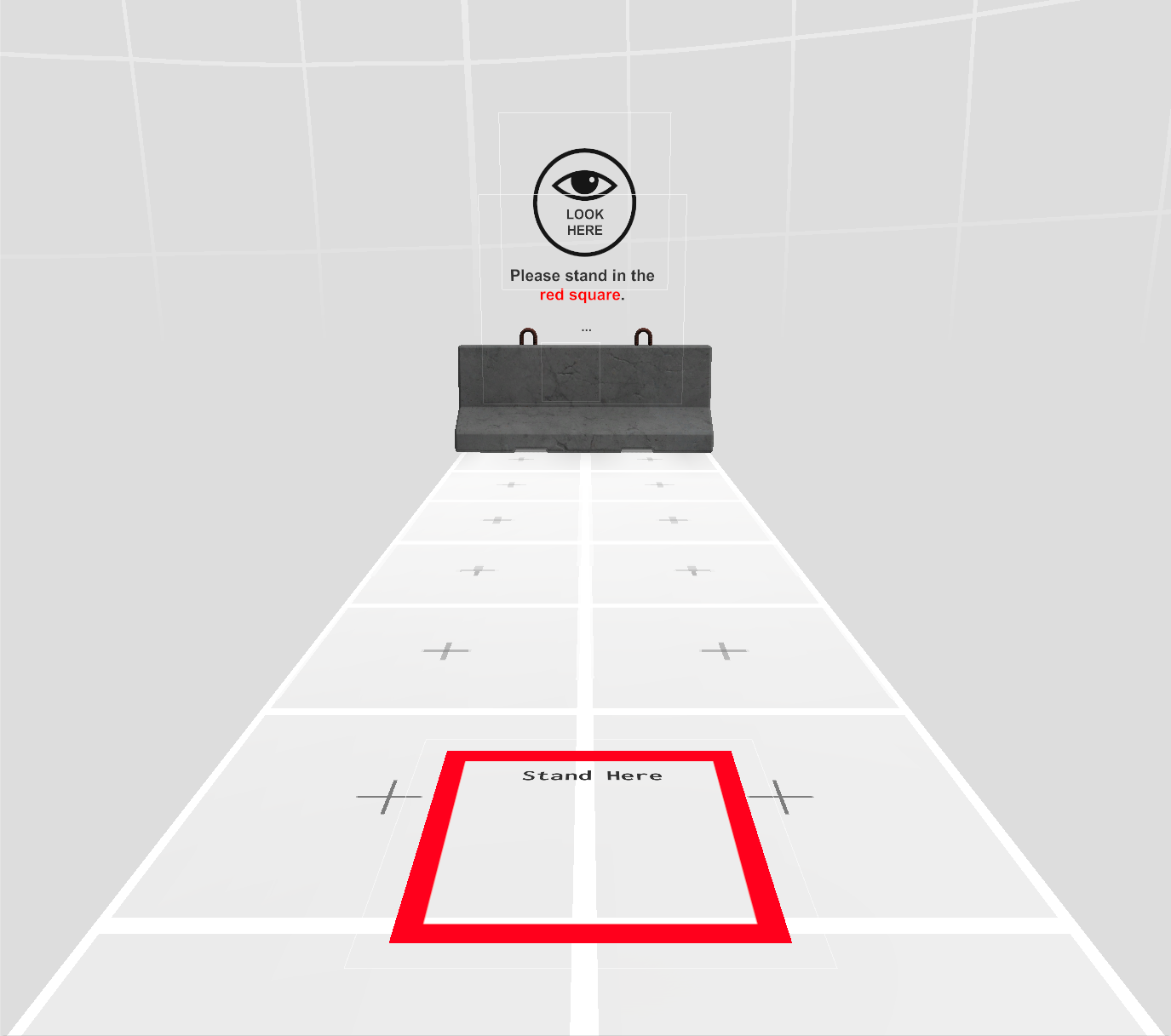}
      \caption{Virtual Staging Environment with Starting Position (red) and View Direction Indicator}
      \label{fig:staging_environment}    
  \end{center}
\end{figure}

Our experiment makes use of a virtual staging environment, depicted in \figref{staging_environment}, where the participants remain between crossing attempts, with a marker for the crossing starting position. During crossing attempts participants are placed in an alleyway, 3 meters wide, 2 meters from the curb of the road. The walls of the alleyway extend up until 0.5 m from the road, preventing the participant from seeing any approaching vehicles until they have stepped out of their initial starting position. The road is 6 meters wide with a continuous yellow lane marking down the middle. \figref{topdown} shows an overview of the virtual environment, \figref{setup} shows the physical setup.  As we only had 6 meters of total physical distance available for both the ally and the road (\figref{setup_illustration} - 3), we returned participants to the staging environment after crossing the first 2.5 m of the first lane (\figref{setup_illustration} - 5), giving them 1.5 meters of buffer space to decelerate. 

\figref{walker} illustrates the interaction between a participant and a virtual test-vehicle via a visual mockup. \figref{fpv} shows how the participant perceives this interaction in the HMD. 

\subsection{Procedure}\label{sec:procedure}

Participants were informed that the intention of the experiment would be to study how the behavior of oncoming vehicles would affect the decisions of pedestrians to cross the road. They were instructed to treat the virtual interactions as they would treat interactions in reality. They were specifically reminded to avoid any risks they would not take with real cars. They were further instructed to act as if in a hurry, to cross ``rather sooner than later'', however not at the risk of bodily harm.

After the instructions the participants put on the Head Mounted Display and were familiarized with the virtual environment. We demonstrated the mechanism which warns VR users when they are about to approach the limits of the VR space and encouraged participants to explore the limits of the virtual environment before beginning the trail. Once they felt comfortable walking inside the environment wearing the HMD, we began the actual study with two introductory interactions. We demonstrated to the participants what would happen if they were to come into contact with the virtual vehicle (an acoustic signal and the immediate return to the staging environment), to discourage them from provoking a ``collision'' out of curiosity. In the second scenario we allowed the participants to cross the street in front of a stopped car to introduce them to the mechanism which would return them to the staging environment after traversing the fist lane.

For each of the crossing attempts, participants would go through the following steps:
\begin{enumerate}
  \item The participant stands in a marked position in the staging environment, gazing at a second marker placed in the direction of our virtual street. 
  \item The scene switches to the street environment, placing the participant in the alley with a limited view of the street.
  \item The participant walks out of an ally and sees a vehicle approach from the left
  \item The moment the car becomes visible to the pedestrian the the trajectory is triggered. 
  Due to this mechanism all participants experience the same \textit{time to arrival (TTA)} for each trajectory. The vehicle approaches the intersection in a straight line in the middle of the lane,  with speed, starting distance and acceleration at any point in time being determined by the trajectory under test in the given attempt.
  \item The participant has to asses whether they want to try to physically walk across the first lane (3 m)
  \item The result and timing of all crossing events is logged automatically. Additionally the participant is requested to provide feedback on a series of questions.
\end{enumerate}


Participants were further asked to answer the following questions after each attempt:
\begin{itemize}
  \item “Describe briefly, what did the car do?” \textit{(open question)}
  \item "Would you say the car was accelerating, decelerating, going at a constant speed or doing something else? \textit{(4\,options)}
  \item “How safe did you feel in this situation?” \textit{(Likert Item)}
  \item “Did the actions of the car surprise you?” \textit{(yes/no)}
  \item “How much trust did you have in this car?” \textit{(Likert Item)}
  \item “Do you believe the car reacted to your presence?” \textit{(yes/no)}
  \item “Would you have acted the same way in the real world?” \textit{(yes/no)}
  \end{itemize}

\subsection{Trajectories}\label{sec:trajectories}

As stated before our crossing scenarios were designed to gauge participant reactions towards different kinds of vehicle behaviors, with the goal to identify a difference in participant reactions between vehicle behaviors designed to comply with social expectations and vehicle behaviors designed to subvert social expectations.

To achieve this, the vehicles in our crossing scenarios followed different \textit{trajectories}. For our purposes a trajectory describes the behavior of an approaching vehicle by determining the vehicle speed and acceleration for any given point in time. Some of these trajectories were interactive, while others were following a predetermined acceleration curve.

For the purpose of the aforementioned comparison we created two distinct groups of trajectories:
\begin{description}
  \item[\tgroup{Yield} (\tgreen):] Trajectories intended to comply with social expectations.
  These trajectories were designed to encourage pedestrians to cross the street. The vehicle slows down aggressively at a certain distance from the pedestrian but keeps rolling at a slow speed in order to elicit a decision for or against crossing.
  
  \item[\tgroup{subversion}: (\tred)] Trajectories in this category were designed with the intention to subvert social expectations. The trajectories display varying degrees of unusual vehicle behaviors, some are just confusing while other are outright malicious. Trajectories in this set are dynamic and react to the actions of the pedestrian, in many cases by accelerating towards them.
\end{description}

In addition to these basic attempts at communication we included two sets of trajectories to study if basic changes in acceleration would yield different reactions. 
Each of these two sets consists of three trajectories with a common final approach velocity and identical TTA. One of the trajectories starts at a lower velocity and accelerates towards the terminal velocity, one trajectory which starts at a higher velocity and decelerates towards the terminal velocity and finally one trajectory with no acceleration change for comparison.

\begin{description}
  \item[\tgroup{15 kph Set} (\tlightblue):] Three trajectories with 15 km/h as the final approach velocity of the vehicle, all with a TTA of 8s. 
  \item[\tgroup{40 kph Set} (\tdarkblue):] Three trajectories with the final approach speed of 40 km/h and a TTA of 8s. 
\end{description}

All trajectories up to this point shared a time to arrival between 8s and 9s, in order to make crossing decisions comparable between them. In addition to these we tested some trajectories with a lower TTA:

\begin{description}
  \item[\tgroup{deterrent} (\tgrey):] Trajectories designed to be challenging to impossible to cross safely, with a time to arrival as low as two seconds. As trajectories from almost all other groups have a TTA of 8s or more or more these are interspersed to prevent participants form believing that crossing the street is possible for all interactions, forcing them to carefully consider the decision to cross each time.
  
  \item[\tgroup{other} (\tpurple):] This group consists only of the trajectory \traj{breaking\_on\_enter}. Vehicles following this trajectory have a comparatively low TTA of 4.8s, but will slow down if the participant steps into the lane of travel. 
\end{description}

Excluding our introductory scenarios we tested a total of 15 trajectories,  The individual trajectories are described in Table \ref{tab:trajectories}. 
Participants completed each trajectory once. The number of trajectories was limited to keep the duration of one session within thirty minutes.

\subsection{Participants}\label{sec:participants}

Participants were recruited from the immediate surroundings of our lab, members of the MIT Center for Transportation and Logistics not involved in the project. 
All participants reported living around the greater Cambridge and Boston area. 
Participants ranged from 22 to 55 years of age, the average age being 32.96, with a standard deviation of 9.15. The total number of participants was 22, 9 female and 13 male. 
Participants were compensated with bananas and donuts. 


\section{Results and Discussion}\label{sec:results}

\paragraph{Road Crossing Decisions}

\begin{figure}
  \includegraphics[width=\columnwidth]{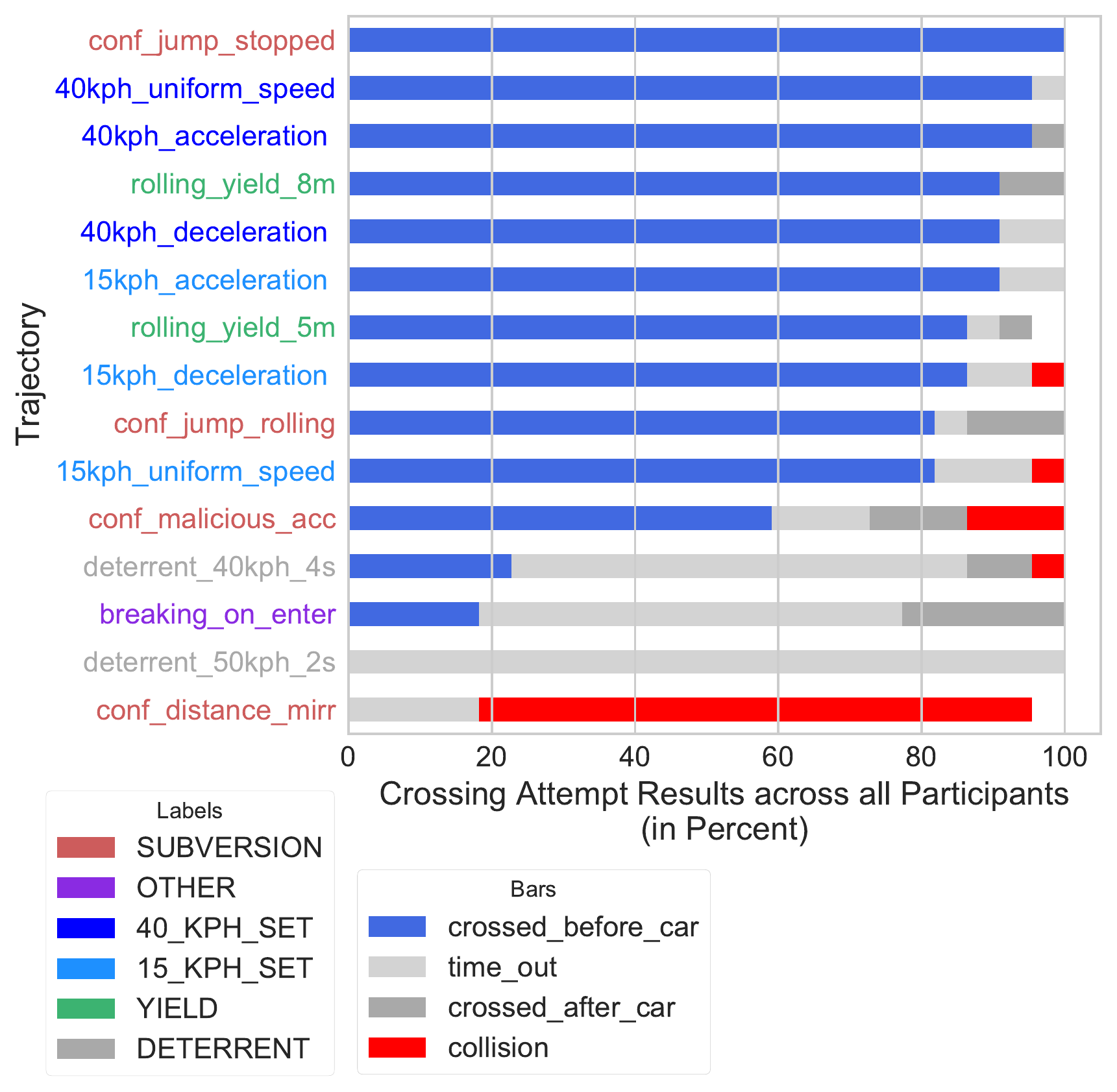}
  \caption{Results of crossing attempts. Label color indicates trajectory group.}
  \label{fig:trajectory_results}
\end{figure}

\begin{figure*}[tp]
  \centering
  \begin{subfigure}[b]{0.49\textwidth}
      \includegraphics[width=\textwidth]{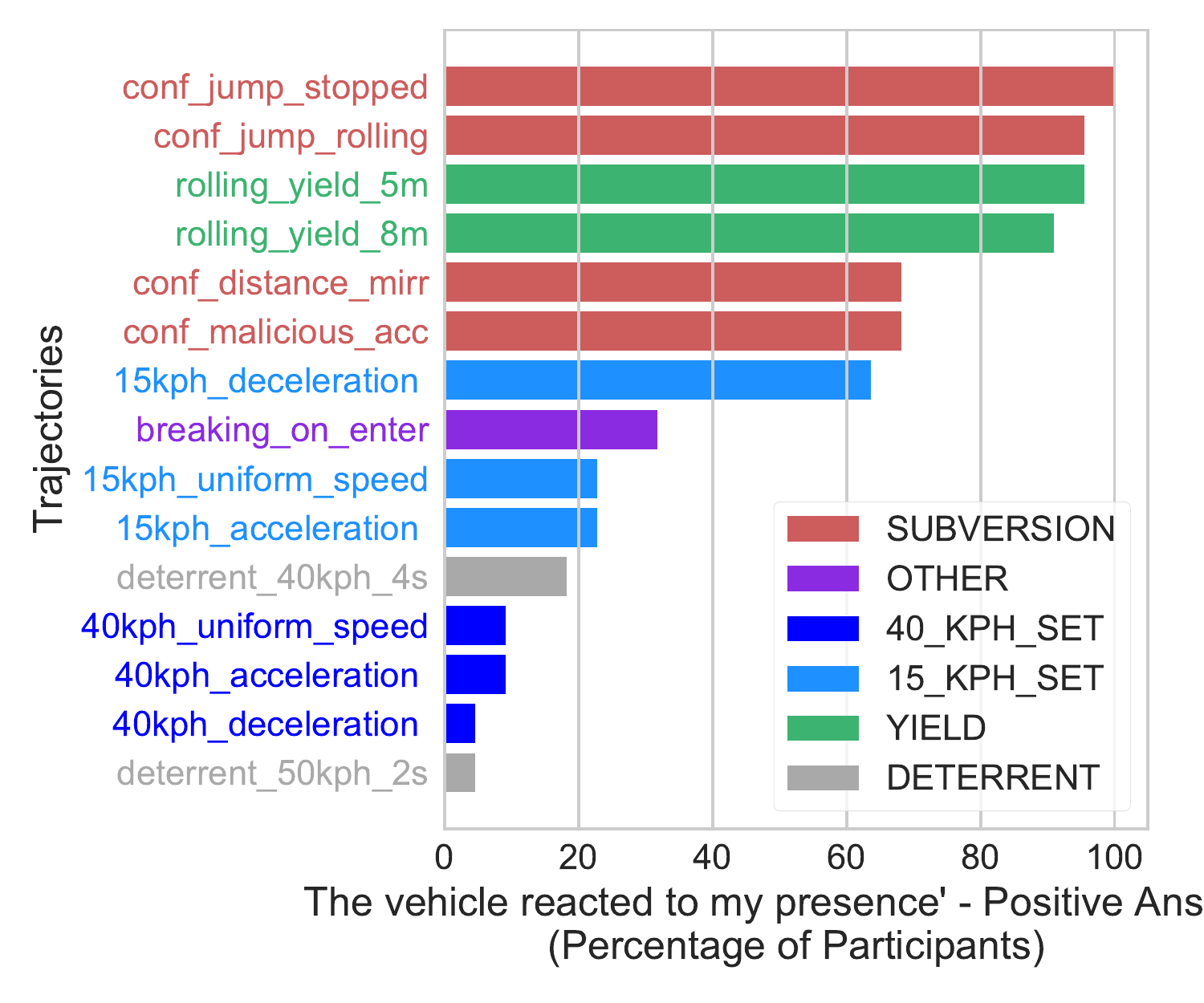}
      \caption{Participants perception of vehicle's reaction to their presence.}
      \label{fig:reacted}
    \end{subfigure}
  \hspace{0.01in}
  \begin{subfigure}[b]{0.49\textwidth}
    \includegraphics[width=\textwidth]{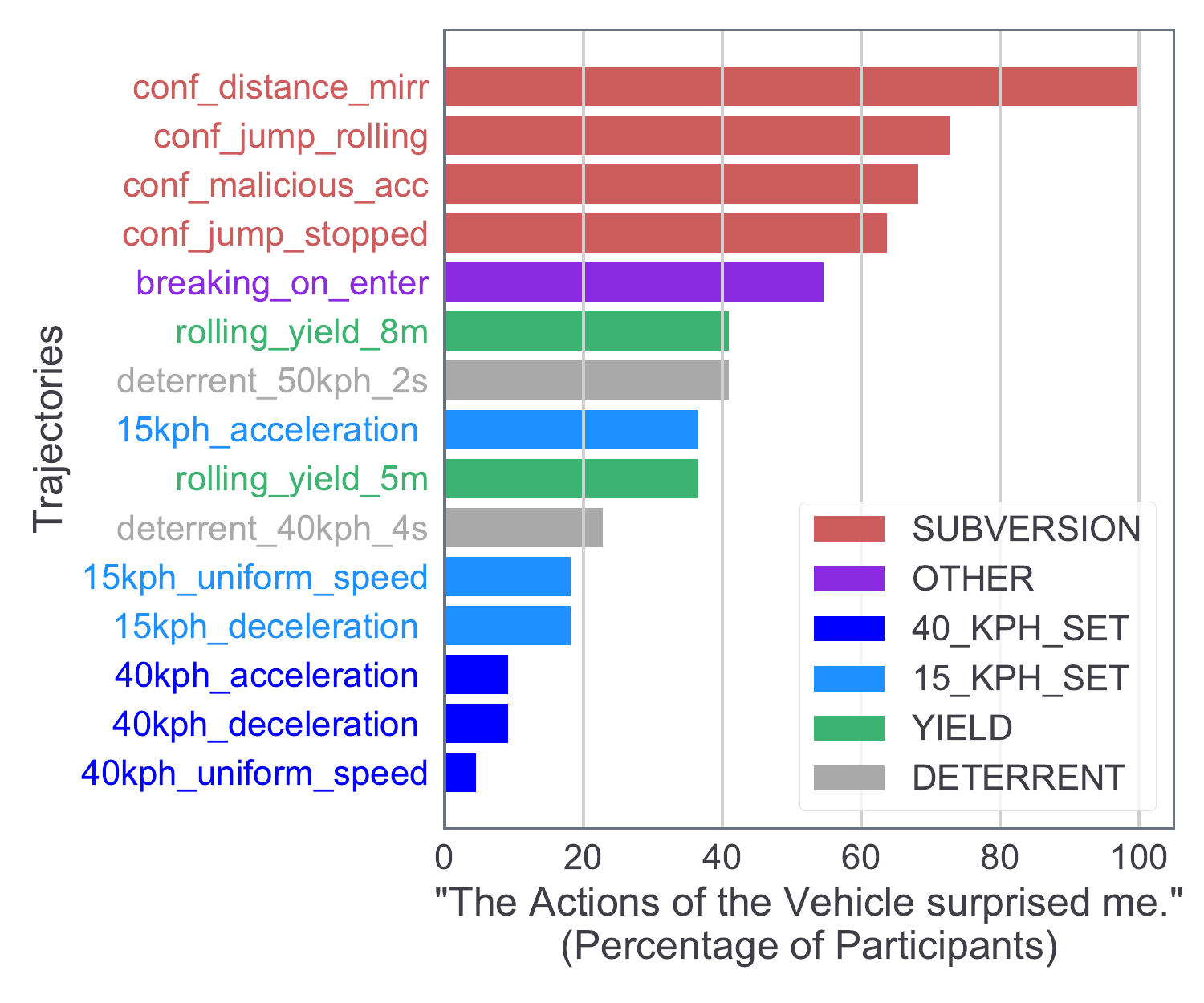}
    \caption{Participants surprised by vehicle behavior.}
    \label{fig:surprise}
  \end{subfigure}
  \caption{Participant reaction towards trajectories.}
  \label{fig:reactions}
\end{figure*}

We recorded a total of 328 individual crossing attempts, excluding two training attempts per participant. 
Two crossing attempts could not be recorded due to technical issues and were excluded from analysis. 

Excluding trajectories from the \tgroup{deterrent} (\tgrey) group as well as the trajectory \traj{conf\_distance\_mirr}, as those trajectories were designed to inhibit road-crossing, that left 263 individual crossing opportunities to study crossing decisions. 
Out of those 263 attempts participants crossed in front of the approaching vehicle 81.75\% of the time. Four of the remaining cases resulted in collisions, the remainder are cases were participants decided not to cross or crossed after the vehicle. 

In the following, ``successful crossing'' will refer to crossing attempts completed by entering the street in front of the approaching vehicle without any collisions.

This high success-rate for crossing opportunities fits the circumstances as for all of these interactions the TTA was 8s and participants were primed to cross if possible. 

It is further consistent with the real-world observations in \cite{rothenbucher2016ghost} where the majority of pedestrians crossed in front of a seemingly autonomous vehicle even if it had shown a transgression towards them during its approach. 

\figref{trajectory_results} provides the success-rate for each trajectory, showing which percentage of participants crossed in front of the approaching vehicle, which percentage crossed after the vehicle had passed (or not at all) and which percentage of participants collided with the vehicle.
Crossing decisions are an important metric given the long-term goal of influencing pedestrian crossing decisions as stated in \ref{sec:introduction}. Furthermore deciding not to cross despite a sufficient gap-distance could be interpreted as a strong signal of a participants reaction to the vehicle behavior in the given trajectory. 

 Looking at trajectories with a lower TTA (see \tabref{trajectories} in \figref{trajectory_results} we can see observations of previous studies regarding crossing decisions hold true in our environment, as these trajectories with a low TTA (five seconds or less), such as the \tgroup{deterrent} (\tgrey) trajectories as well as \traj{braking\_on\_enter} show the least amount of
crossings completed successfully. This is an argument towards the perceived
realism of our simulation.

\traj{con\_distance\_mirr} has a high number of ``collisions'' as this trajectory did not offer any other solution to the scenario except waiting for the time limit to pass. 





\paragraph{Reacting to Presence} 

\begin{figure*}
  \centering
    \centering
    \includegraphics[width=\textwidth]{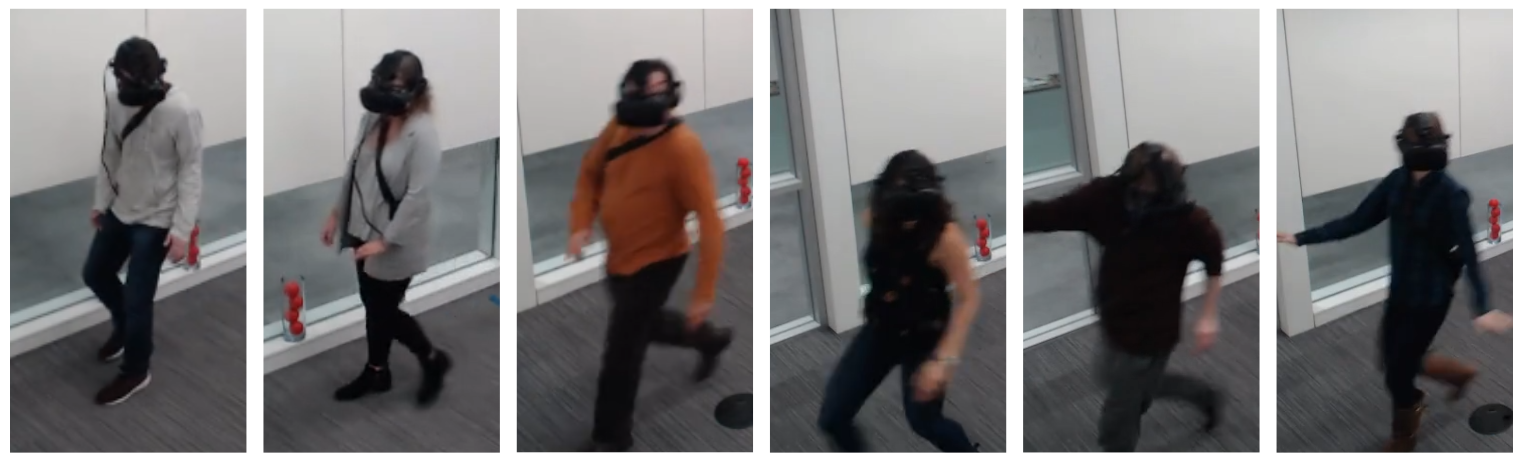}
    \captionsetup{width=0.9\linewidth}
    \captionof{figure}{Participants Reacting Strongly To Transgressions From The Simulated Vehicle}
    \label{fig:transgression}
\end{figure*}

Given the overall goal of using vehicle kinematics as a means for communicating with pedestrians it is important that pedestrians perceive actions taken by the vehicle as a reaction to their presence, otherwise communication cannot occur, at least on a conscious level. 

\figref{reacted} shows which percentage of participants believed the actions of the vehicle were a reaction to their presence for each trajectory. This was self reported by participants after each crossing attempt.

It can be observed that the trajectories belonging to the two sets designed to communicate with pedestrians, the \tgroup{subversion} (\tred) set as well as the \tgroup{yield} (\tgreen) set, were indeed perceived as interactive by the largest percentage of participants.  Furthermore we see that trajectories designed without the intention to communicate, such as the \tgroup{DETERRENT} (\tgrey) trajectories as well as a trajectories featuring a ``uniform speed'' rank a lot lower in comparison. 

This strongly supports the possibility that trajectories can be used to intentionally convey information. 

Looking closer at the four trajectories belonging to the \tgroup{SUBVERSION} (\tred) set, we a difference between the trajectories meant to be irritating \traj{conf\_jump\_stopped}, \traj{conf\_jump\_stopped} and the hostile trajectories \traj{conf\_distance\_mirr} and \traj{conf\_malicious\_acc}, with the latter ones ranking lower in perceived interactivity. This is consistent with comments made by some participants who did not consider malicious behavior to be a possibility, providing statements such as \textit{``The fact that it accelerated into my path made me believe that was [originally] stopping for a factor that was not me''} (\traj{conf\_distance\_mirr}). Instead, such behavior was often attributed to negligence.
In terms of breaking social conventions this would imply that the malicious behavior is so far removed from the expected norm that it is not even considered as a possibility for these interactions, which points towards the existence of a social norm. 


\paragraph{Subverted Expectations} \label{par:subverted_expectations}

To determine if we succeeded in subverting the expectations of street-crossing interactions we queried our participants after every attempt if they were surprised by the behavior of the vehicle. \figref{surprise} shows for each trajectory which percentage of participants were surprised by the actions of the vehicle.

The trajectories from the \tgroup{SUBVERSION} (\tred) set were perceived as surprising by a greater
percentage of participants than all other trajectories. I can therefore be stated that the \tgroup{SUBVERSION} (\tred) trajectories succeeded in their design goal of subverting pedestrian expectations, which in combination with the participant feedback we received suggests a social component in the interpretation of vehicle kinematics exists.

\traj{15\_kph\_acceleration} was perceived as surprising by twice as many participants than the other two trajectories from the same set (\tgroup{15\_KPH\_SET}, \tlightblue), suggesting that accelerating in the presence of pedestrians might be considered to be outside of the social norm, however multiple participants also cited the slow initial speed of the vehicle as being unusual and the reason for their confusion.


\begin{figure*}
  \centering
  \begin{minipage}[t]{.5\textwidth}
    \centering
    \includegraphics[width=\textwidth]{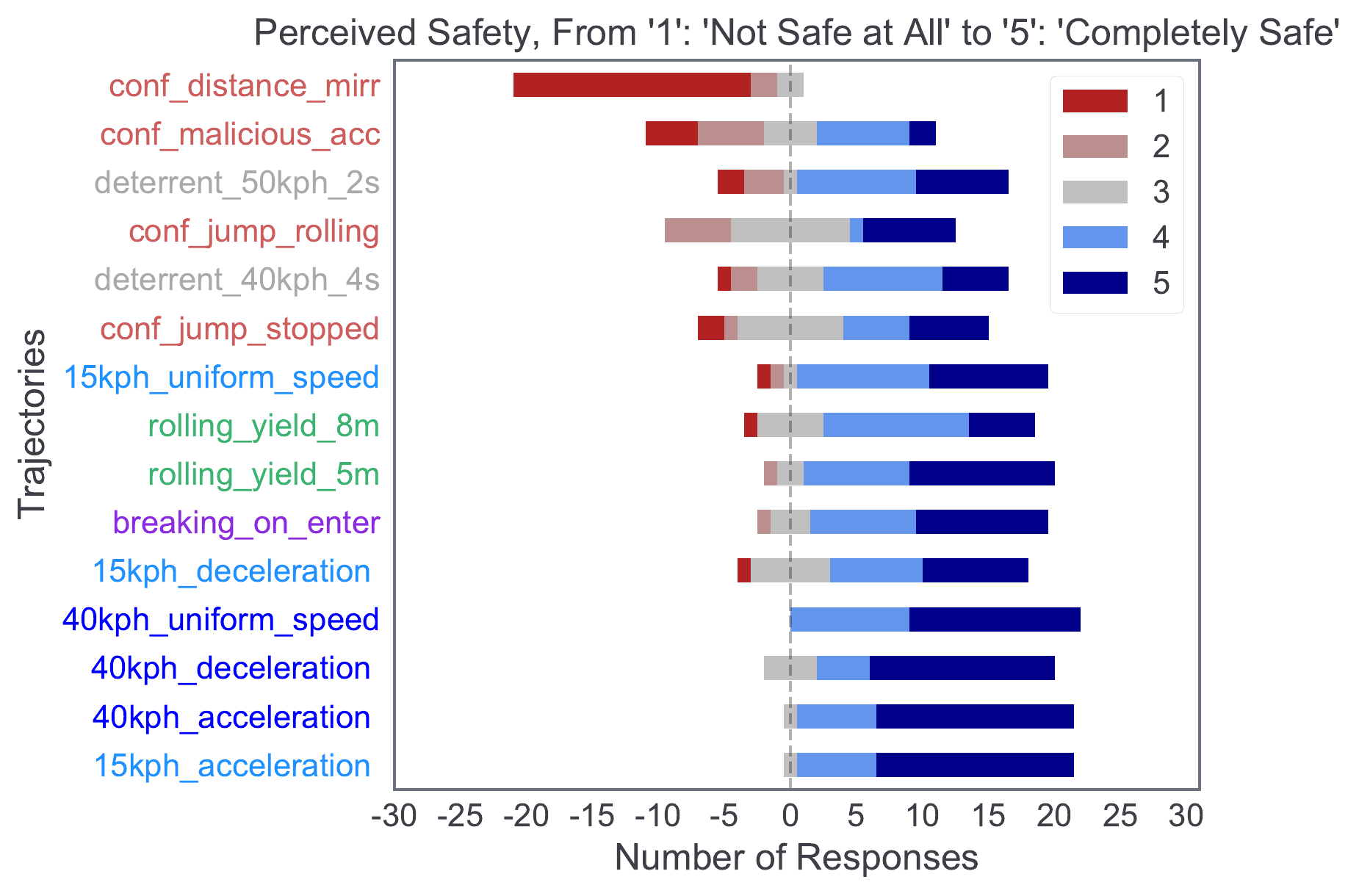}
    \captionsetup{width=0.9\linewidth}
    \captionof{figure}[3in]{Perceived safety during Interaction. Label color indicates trajectory group.}
    \label{fig:safety}
  \end{minipage}%
  \begin{minipage}[t]{.5\textwidth}
    \centering
    \includegraphics[width=\textwidth]{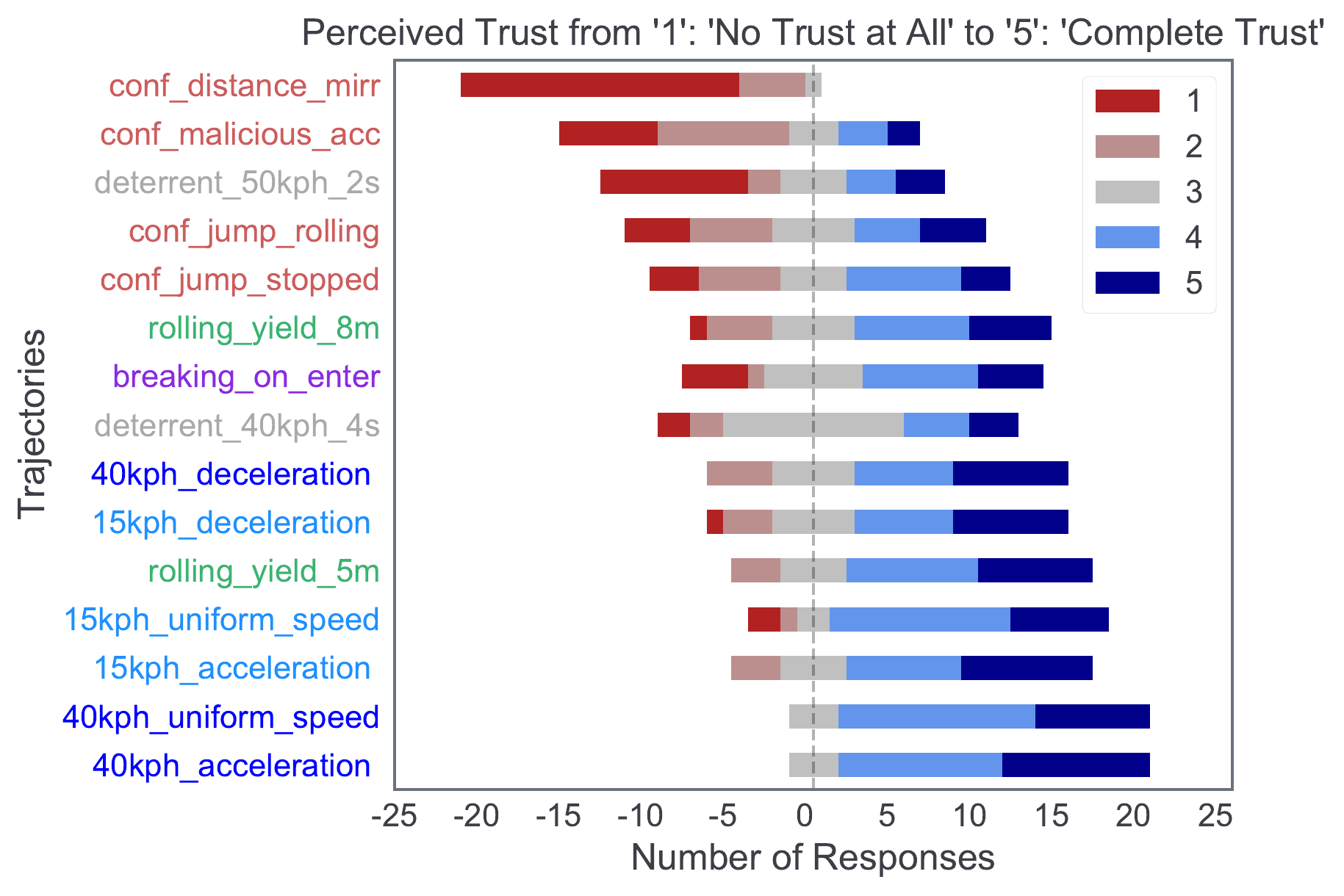}
    \captionsetup{width=0.9\linewidth}
    \captionof{figure}{Trust in vehicle trajectory. Label color indicates trajectory group.}
    \label{fig:trust}
  \end{minipage}
\end{figure*}

\paragraph{Interpreting the Vehicle Behavior in a Social Context} \label{par:social_context}

In searching for a social context the interpretation of vehicle kinematics the open feedback provided by participants was very instructive.

Looking specifically at subversive trajectories of \traj{conf\_jump\_stopped} and \traj{conf\_jump\_moving}, we observed two different interpretations of the vehicle behavior by the participants.
For both trajectories the car is either stopped or moving very slowly and then accelerates briefly when a participant approaches the curb while looking at the vehicle, before returning to the initial speed. Depending on the behavior of the participant this can be repeated multiple times (see also \tabref{trajectories}).

Our participants were split in their interpretation of this behavior: 

The first group of participants believed the vehicle started to accelerate as a reaction to their presence, which is in accordance with the design of the trajectory. Some of these participants were perplexed by this behavior as we had intended, with their evaluation of the situation ranging from \textit{``they were kind of being annoying''}, \textit{``so weird''}, \textit{``unclear''} and \textit{``unpredictable''} to \textit{``it was intentionally trying to make me scared''}. 

The second group of participants assumed that the acceleration of the vehicle happened because the vehicle was not aware of their presence (\textit{``a failure of attention''}), while the deceleration was seen as a reaction to the vehicle registering their presence, with one participant explaining: \textit{``[I felt] high trust, because [the car] immediately braked when it saw me''}.

The following statements were of particular interest:

\begin{itemize}
  \item \textit{``Call me paranoid, but the way it stopped I wasn't sure it wasn't going to accelerate as I started to cross.''}
  \item \textit{``He first accelerated like he wanted to be first but then stopped.''}
  \item \textit{``[It] felt like it was trying to intimidate me or something [it then stopped] to let me go, after he thought about possibly not letting me go.''} 
  \item \textit{``It appeared the driver was not sure if they wanted to let me go or not''}
  \item \textit{``It was a \textbf{social thing} - you go, no, you go''}  
  \item \textit{``The driver clearly saw me, but he did not see me right away so I did not know how much attention he was paying to me.''}
  \item \textit{``One strike for not seeing me in the beginning, but it then compensated for that by stopping.''}
\end{itemize}

It is important to emphasize that our questions always explicitly referred to ``the car'', meaning any mention of a
driver as well as personifications using 'he' are an unprompted choice by the individual participants. The previous
exemplary statements not only show that the participants perceived a social component in the trajectory, they
were reflecting on the intentions of the vehicle as an entity in the context of their own actions and intentions. These responses strongly support the surprise metric as seen in \figref{surprise}.

We believe the comments given are another strong indication of the sense of presence experienced by the participants in the simulation. This is further supported by the fact that some participants were gesturing towards the virtual vehicle and reacted very strongly towards the ``physical'' presence of the vehicle, especially during the transgressions induced by the subversive trajectories as can be seen in \figref{transgression}.




\paragraph{Trust and Safety}

Our expectation was that the adherence of vehicles to a potential social construct would affect the predictability of their behavior and by extension the how for pedestrians trust them in an interaction. 

\figref{trust} shows a Likert-item rating of the trust participants felt towards vehicles following the different trajectories, on a scale of ``1'' - ``no trust at all'' to  ``5' - ``complete trust''. 

The trajectories in \figref{trust} are ordered based on the total number of Likert-item responses given which are less than ``three''.

We can see that the trajectories designed to subvert social expectations, \tgroup{subversion} (\tred) and discourage crossing all together \tgroup{deterrent} (\tgrey), did in fact receive the lowest trust-ratings from our participants. The trajectory mirroring pedestrian behavior, \traj{conf\_distance\_mirr} received a distinctly negative rating with the highest number of ``no trust at all'' ratings out of all trajectories. Several participants describe the car as \textit{``playing a game''}, with one person labeling the vehicle as a \textit{``psychopath.''}

We can see that all of our subversive trajectories were in fact perceived as irritating. Since we prompted our participants ``to cross if possible'' as if they were in a hurry, it is hard to tell if under other conditions the diminished trust in the vehicle would have let to fewer decisions to cross in front of it. Looking back at \figref{trajectory_results} that a majority of pedestrians still crossed despite feeling uneasy in the case of the \tgroup{subversion} (\tred) trajectories is an interesting observation. In any case, the lower rating of trust compared to other trajectories not designed to subvert social expectations, such as the \tgroup{yield} (\tgreen) trajectories further supports the notion that vehicle kinematics are used to judge the the these interactions with a social component. 


Besides asking about trust, we also asked participants to rate their feeling of safety in the interactions as a Likert-item (\figref{safety}). We were interested if the unpredictable nature of some of our trajectories would affect how safe participants would feel in these interactions.

With reference to \figref{safety} we can see that participants reported feeling safe for a great majority of the interactions. This is not particularly surprising as participants were instructed not attempt a road crossing if the situation could result in injury if were to happen outside of our simulation. 
Nevertheless it is interesting that one of our openly malicious trajectories, \traj{conf\_distance\_mirr} only received ratings below a neutral ``3''.


\paragraph{Judging Acceleration}

The ability to communicate by means of vehicle kinematics requires the ability in pedestrians to perceive and identify how the vehicle moves, particularly if it is changing its velocity.

\begin{figure} 
  \begin{center}
      \includegraphics[width=\columnwidth]{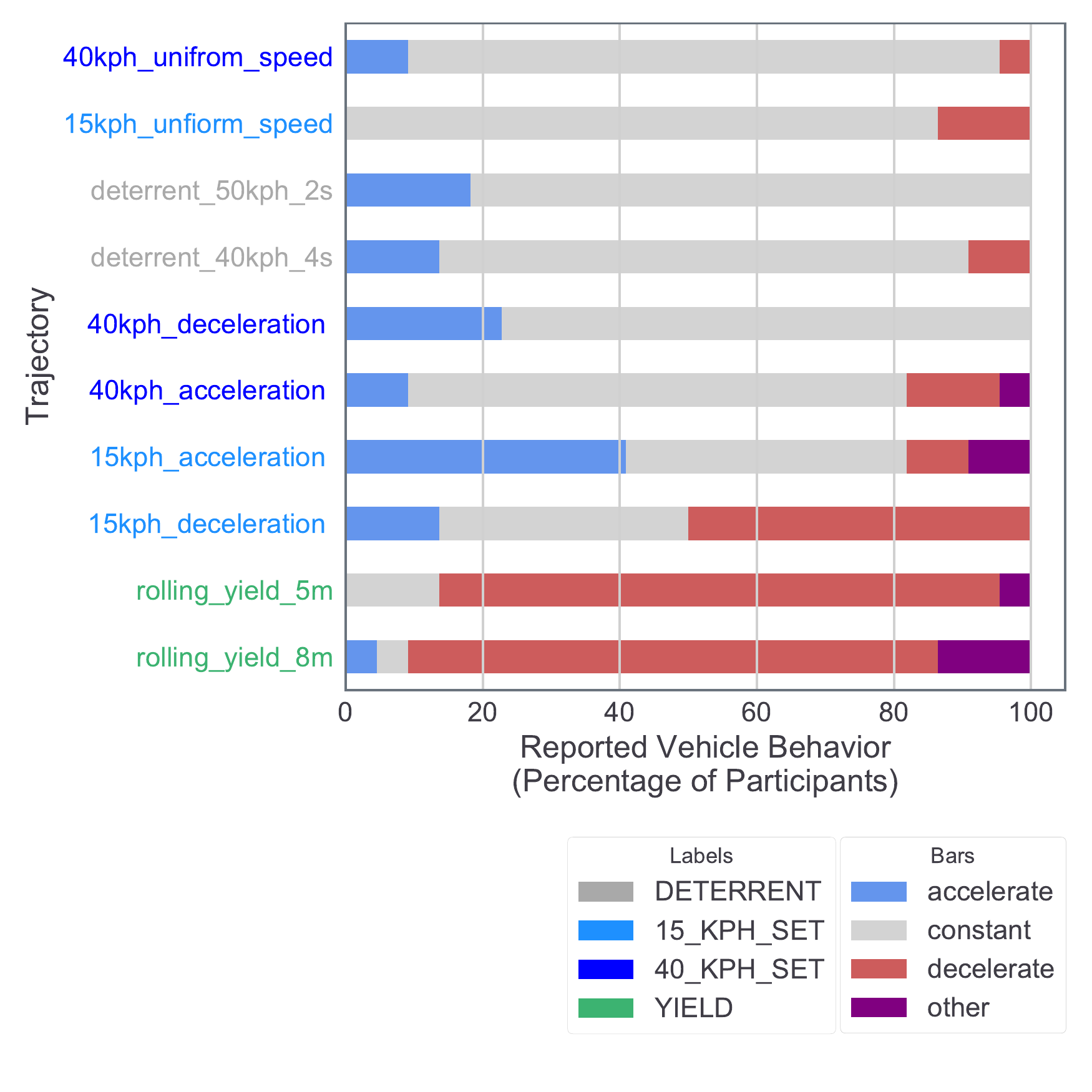}
      \caption{Vehicle acceleration behavior as observed by participants.}
      \label{fig:guess_answers}    
  \end{center}
\end{figure}

To test this ability, we queried our participants after each interaction to sort the movement into one of four categories:  ``accelerating'', ``decelerating'' or ``going at a constant speed''. 
\figref{guess_answers} shows how the responses per category given by participants for each trajectory. The figure features only those trajectories containing a single acceleration change, as labeling multiple sequential changes as occurred in some of the interactive trajectories would be significantly more complicated.

It can be observed that at higher speeds and greater distances a majority of participants default to ``constant speed'' independent of the presence of acceleration changes in the trajectory as those changes become harder to observe, while at lower speeds with  the \tgroup{YIELD} (\tgreen) set and the \tgroup{15\_kph\_set} (\tlightblue) the majority of participants identify acceleration and deceleration correctly. 

The limit of such perception poses a limit on the situations in which communication via kinematics could be applied and requires further study.



\section{Conclusion}\label{sec:conclusion}

Our goal was to study if pedestrians derive social clues from vehicle kinematics, if such interactions could be studied in virtual reality and to estimate the potential in using vehicle kinematics for effective communication in autonomous vehicles.

We confronted our participants with different vehicle kinematics, some of witch were designed to subvert social expectations while others were intended to be conform with expectations. We were able to show that our participants perceived the changes in vehicle motion as a direct reaction to their presence. We were able to show that vehicles following intentionally atypical trajectories let to confusion and in some cases mistrust among participants, while more conventional trajectories did not.

Previously vehicle kinematics in the context of pedestrian interactions have been viewed as a matter of physics, with pedestrians assessing if the approaching vehicle leaves them enough time to cross its path of travel (evaluation of gap distance).

The data we collected and the remarks we received from our participants show, that pedestrians evaluate vehicle kinematics beyond a consideration of time to arrival, as a social interaction from which they derive cues, going so far as to reflecting on the driving entities perception of their own intentions. 

We were able to make these observations in an immersive virtual reality simulation, which leads us to conclude that VR is a valid tool for further exploration of this concept. 

We believe that future work will enable the use of vehicle kinematics to communicate driving intentions to pedestrians.




\section*{Acknowledgment} 

This work was in part supported by the Toyota Collaborative Safety Research Center. The views and conclusions being
expressed are those of the authors and do no necessarily reflect those of Toyota.

\balance

\bibliography{vr}



\end{document}